\documentclass{ws-mpla}
\usepackage{amsmath}
\usepackage{amssymb}
\usepackage{graphicx}

\newcommand{\pat}{\partial}

\begin{document}

\markboth{A.V.Minkevich} {On theory of regular accelerating Universe in
Riemann-Cartan spacetime}

%

\title{On theory of regular accelerating Universe in Riemann-Cartan spacetime}

\author{A.V. Minkevich}
 \address{Department of Theoretical Physics and Astrophysics, Belarussian State University,\\
 Minsk, Belarus\\
minkav@bsu.by}
  \address{Department of Physics and Computer Methods, Warmia and Mazury University in
 Olsztyn,\\
 Olzstyn, Poland\\
 awm@matman.uwm.edu.pl}

\maketitle


\begin{abstract}
Isotropic cosmology built in the Riemann-Cartan spacetime is investigated.
Properties of homogeneous isotropic cosmological models filled with usual
gravitating matter and scalar fields are studied in the beginning of
cosmological expansion near the limiting energy density. It is shown that
cosmological models are regular not only with respect to the Hubble parameter
and the energy density but also with respect to the torsion and curvature
tensors.

\end{abstract}

\ccode{PACS numbers: 04.50.+h; 98.80.Cq; 11.15.-q; 95.36.+x}

\section{Introduction}

The problem of the beginning of the Universe in time -- the problem of
cosmological singularity (PCS) -- and the problem of invisible components of
gravitating matter, dark energy and dark matter, remain as the most principal
problems of relativistic cosmology built on the base of general relativity
theory (GR). Many attempts were undertaken with the purpose to solve these
problems in the framework of GR and existent candidates to quantum gravitation
theory (string theory/M-theory, loop quantum gravity) as well as different
generalizations of GR. Radical ideas connected with the notions of strings,
additional spacetime dimensions, spacetime quantization etc. were used. Various
hypothetical fields and particles with unusual properties as possible
candidates to dark energy and dark matter were discussed in literature. It
should be noted that many existent generalizations of Einsteinian gravitation
theory are based on ad hoc hypotheses and do not have solid theoretical
foundation.

At the same time there is the gravitation theory built on the base of generally
accepted field-theoretical principles including the local gauge invariance
principle which is a natural generalization of GR and which offers
opportunities to solve its principal problems: the Poincar\'e gauge theory of
gravity (PGTG) -- the gravitation theory in 4-dimensional physical spacetime
with the structure of Riemann-Cartan continuum $U_4$. The formation of PGTG is
inseparably connected with works \cite{al,a2}, from which follows that
gravitation theory in Riemann-Cartan spacetime is a necessary generalization of
GR by including the Lorentz group (the group of tetrad Lorentz transformations)
to gauge group corresponding to gravitational interaction \footnote{The
bibliography of works dedicated to PGTG is given in \cite{a3}.}. The simplest
PGTG is Einstein-Cartan theory of gravity based on gravitational Lagrangian in
the form of scalar curvature of $U_4$ \cite{a4}. In the case of spinless matter
gravitational equations of Einstein-Cartan theory are identical to Einstein
gravitational equations of GR, and in the case of spinning matter the
Einstein-Cartan theory leads to linear relation between spacetime torsion and
spin momentum of gravitating matter. Because of this fact the opinion that the
torsion is generated by spin momentum of gravitating matter is widely held in
literature. However, by taking into account that the torsion tensor plays the
role of gravitational field strength corresponding to subgroup of spacetime
translations connected directly with energy-momentum tensor, we can conclude
that this fact discusses the degenerate character of Einstein-Cartan theory.
The situation comes to normal by including to gravitational Lagrangian
similarly to the theory of Yang-Mills fields terms quadratic in gauge
gravitational field strengths - the curvature and torsion tensors \cite{a5,a6}.

By using sufficiently general expression of gravitational Lagrangian of PGTG
including both a scalar curvature and quadratic terms in curvature
($F_{\alpha\beta\mu\nu}$) and torsion ($S_{\alpha\mu\nu}$) tensors with
indefinite parameters isotropic cosmology was built and investigated in a
number of papers (see Refs. 7-15 and references herein). It was shown that
gravitational interaction in the frame of homogeneous isotropic models (HIM) by
certain restrictions on indefinite parameters is changed in comparison with GR
and can be repulsive by certain conditions that allow solving the PCS and also
to explain the acceleration of cosmological expansion at present epoch without
using the notion of dark energy. From cosmological equations deduced in
Ref.\cite{a14} follows that gravitational repulsion effect takes place at
extreme conditions (extremely high energy densities and pressures) at the
beginning of cosmological expansion by virtue of existence of limiting
(maximum) energy density. As a result all cosmological solutions for HIM filled
with gravitating matter satisfying standard energy conditions are regular, and
they contain the compression stage before cosmological expansion stage.
Gravitational repulsion effect appears also when energy density in HIM is very
small that leads to cosmological solutions for accelerating Universe. Repulsion
effect in this case has the vacuum origin because the physical spacetime in the
vacuum has the structure of Riemann-Cartan continuum with de Sitter metrics and
non-vanishing torsion \cite{a13}. The change of gravitational interaction in
the frame of PGTG in comparison with GR is connected with more complicated
structure of physical spacetime, namely with spacetime torsion.

The present paper is devoted to analysis of properties of HIM for accelerating
Universe at extreme conditions in the beginning of cosmological expansion. The
principal relations of isotropic cosmology built in the frame of PGTG are given
in Section 2.

\section{Principal Relations of Isotropic Cosmology in PGTG}

In the frame of PGTG any HIM is described by means of three functions of time:
the scale factor of Robertson-Walker metrics $R$ and two torsion functions
$S_1$ and $S_2$ determining non-vanishing components of torsion tensor
$S^\lambda{}_{\mu\nu}=-S^\lambda{}_{\nu\mu}$ \cite{a5}:
$S^1{}_{10}=S^2{}_{20}=S^3{}_{30}=S_{1}(t)$, $\displaystyle S_{123} = S_{231} =
S_{312} = S_{2}(t)\frac{R^3r^2}{\sqrt{1-kr^2}} \sin{\theta}$, where spatial
spherical coordinates are used. Then non-vanishing components of curvature
tensor are determined by means of the following functions $A_i$ ($i=1,2,3,4$):
\begin{eqnarray}\label{1}
&&
    A_1=\dot{H}-2\dot{S}_1+H(H-2S_1),
    \nonumber\\
&&    A_{2}  = \frac{k} {{R^2 }} + \left( {H - 2S_1 } \right)^2  - S_2^2,
    \nonumber\\
&&    A_{3}  = 2\left( {H - 2S_1 } \right)S_2,
    \nonumber\\
&&    A_{4}  = \dot S_2+HS_2,\nonumber 
\end{eqnarray}
where $H=\dot{R}/R $ is the Hubble parameter and a dot denotes the
differentiation with respect to time.

Isotropic cosmology was built by using the following expression of
gravitational Lagrangian (definitions and notations of \cite{a10} are used
below):
\begin{eqnarray} \label{1}
 {\cal L}_{\rm g}=f_0\,
F+F^{\alpha\beta\mu\nu}(f_1\:F_{\alpha\beta\mu\nu}
                +f_2\:
                F_{\alpha\mu\beta\nu}+f_3\:F_{\mu\nu\alpha\beta})
        + F^{\mu\nu}( f_4\:F_{\mu\nu} \nonumber
 \\ + f_5\: F_{\nu\mu}) +  f_6\:F^2
        +S^{\alpha\mu\nu} (a_1\:S_{\alpha\mu\nu}+a_2\:
        S_{\nu\mu\alpha})
    +a_3\:S^\alpha{}_{\mu\alpha}S_\beta{}^{\mu\beta}.
\end{eqnarray}

The Lagrangian ${\cal L}_{\rm g}$ includes the parameter $f_0=(16\pi G)^{-1}$
($G$ is Newton's gravitational constant, the light velocity $c=1$) and a number
of indefinite parameters: $f_i$ ($i=1,2,...6$) and $a_k$ ($k=1,2,3$).
Gravitational equations of PGTG corresponding to gravitational Lagrangian
${\cal L}_{\rm g}$ allow one to obtain cosmological equations generalizing
Friedmann cosmological equations of GR and equations for torsion functions
given in general form in Ref. 13. These equations contain five indefinite
parameters:
\begin{eqnarray}
  a = 2a_1  + a_2  + 3a_3, \qquad b = a_2  - a_1,
\nonumber\\
  f = f_1  + \frac{{f_2 }} {2} + f_3  + f_4  + f_5  + 3f_{6}\, ,
\nonumber\\
  q_1  = f_2  - 2f_3  + f_4  + f_5  + 6f_{6}, \qquad q_2  = 2f_1  - f_2,
\nonumber
\end{eqnarray}
and their mathematical structure and physical consequences depend essentially
on restrictions on these parameters. As it was discussed in Ref. 14, the
simplest HIM possessing important physical properties take place if $a=0$ and
$q_2=0$. The first restriction $a=0$ leads to excluding of higher derivatives
of the scale factor $R$ from cosmological equations \cite{a5}, and the
appearance of limiting energy density in the case of HIM with two torsion
functions is connected directly with the second restriction $q_2=0$ \cite{a14}.
We also give the main relations of isotropic cosmology by using these
restrictions on indefinite parameters.

Cosmological equations generalizing Friedmann cosmological equations of GR take
the following form:
\begin{eqnarray}\label{1}
    \frac{k}{R^2} + (H-2S_1)^2 -S_2^2=
    \frac{1}{{6f_0 Z}}
        \left[
            {\rho  -6 b S_2^2
            + \frac{\alpha }{4} \left( {\rho  - 3p - 12bS_2^2 } \right)^2 }
        \right],
\end{eqnarray}
\begin{eqnarray}\label{2}
    \dot{H}-2\dot{S}_1+H(H-2S_1) =
    -\frac{1} {{12f_0 Z}}
        \left[
            \rho  + 3p - \frac{\alpha } {2} \left( {\rho  - 3p - 12bS_2^2 } \right)^2
        \right],
\end{eqnarray}
where $\rho$ is the energy density, $p$ is the pressure, the parameter
$\alpha=\frac {f} {3f_0^2}$ ($f>0$) has inverse dimension of energy density and
$Z=1+\alpha\left( \rho - 3p - 12b S_2^2\right)$. Cosmological equations (2)-(3)
determine the curvature functions $A_1$ and $A_2$ as functions of matter
parameters $\rho$ and $p$ and of the torsion function $S_2$ obtained from
gravitational equations for HIM. The torsion function $S_1$ is determined by
the following way:
\begin{eqnarray}\label{4}
    S_1  = -\frac{\alpha }{4Z} [\dot \rho
    - 3 \dot p + 12f_0 \omega H S_2^2
    -12( {2b - \omega f_0 } ) S_2 \dot S_2],
\end{eqnarray}
where dimensionless parameter $\omega= \frac {2f -q_1} {f} \neq 0$ is
introduced. The torsion function $S_2^2$ depends on energy density and pressure
as
\begin{eqnarray}\label{3}
 S_{2}^{2}  = \frac{\rho - 3p}{12b} + \frac
{1-(b/2f_0) (1 +  \sqrt{X})} {12b \alpha (1- \omega/4)},
\end{eqnarray}
where $X=1+ \omega (f_0^2/b^2) [1- (b/f_0) - 2(1- \omega /4) \alpha ( \rho +
3p)]$. In order to investigate inflationary cosmological models, we will
consider HIM filled besides usual gravitating matter with energy density
$\rho_m >0$ and pressure $p_m \ge 0$ also by scalar field $\phi$ with potential
$V=V(\phi)$. By neglecting the interaction between these two components of
gravitating matter, the total energy density $\rho$ and pressure $p$ are as
follows:
\begin{eqnarray}\label{6}
\rho=\frac{1}{2}\dot{\phi}^2+V+\rho_m \quad (\rho>0), & &
p=\frac{1}{2}\dot{\phi}^2-V+p_m.
\end{eqnarray}
By minimal coupling with gravitation the equations for gravitating matter take
the usual form as in GR:
\begin{equation}\label{7}
\dot{\rho}_m+3H\left(\rho_m+ p_m\right)=0,
\end{equation}
\begin{equation}\label{8}
\ddot{\phi}+3H\dot{\phi}=-\frac{\partial V}{\partial \phi}.
\end{equation}
By certain restrictions on indefinite parameters cosmological equations (2)-(3)
take at asymptotics, when energy density is sufficiently small, the form of
Friedmann cosmological equations of GR with effective cosmological constant
induced by the torsion function $S_2$ and describe accelerating Universe in
accordance with standard $\Lambda CDM$-model of GR \cite{a14}. In order to
investigate HIM at extreme conditions in the beginning of cosmological
expansion, we transform Eqs. (2)-(8) to dimensionless form by introducing
dimensionless units for all variables and parameter $b$ entering these
equations and denoted by means of \ $\tilde{}$\ as follows:
\begin{equation}\label{dimless}
\begin{array}{lcl}
 t\to\tilde{t}=t/\sqrt{6 f_0 \omega\alpha},& {}
            & R\to\tilde{R}=R/\sqrt{6f_0 \omega\alpha},\\
    \rho\to\tilde{\rho}=\omega\alpha\,\rho, & & p\to\tilde{p}=\omega\alpha\,p,\\
    \phi\to \tilde{\phi} = \phi/\sqrt{6f_0}, & &
            b\to\tilde{b} = b/f_0, \\
    H\to\tilde{H}=\tilde{R}'/\tilde{R}=H\sqrt{6f_0 \omega\alpha}, & &
    S_{1,2}\to\tilde{S}_{1,2}=S_{1,2}\sqrt{6f_0 \omega\alpha},
\end{array}
\end{equation}
where the differentiation with respect to dimensionless time $\tilde{t}$ is
denoted by means of the prime. Obviously Eqs. (6)-(8) conserve their form by
this transformation, namely
\begin{eqnarray}
\tilde{\rho}=\frac{1}{2} {\tilde{\phi}}'^2+\tilde{V}+\tilde{\rho}_m, & {} &
\tilde{p}=\frac{1}{2}{\tilde{\phi}}'^2-\tilde{V}+\tilde{p}_m,
\nonumber\\
 \tilde{\rho}_m'+3\tilde{H}\left(\tilde{\rho}_m+\tilde{p}_m \right)=0,
 & {} & \tilde{\phi}''+3\tilde{H}\tilde{\phi}'=-\frac{\pat
\tilde{V}}{\pat\tilde{\phi}}.
\end{eqnarray}
The transformation of the torsion function $S_2$ defined by (5) to
dimensionless form according to (9) gives:
\begin{eqnarray}\label{11}
 \tilde {S}_{2}^2 =\frac{\tilde{\rho} - 3\tilde{p}}{2\tilde{b}} +
\omega \frac {1-(\tilde{b}/2) (1 +  \sqrt{X})} {2\tilde{b} (1- \omega/4)} ,
 \end{eqnarray}
where
\begin{eqnarray}\label{12}
X=1+ \frac {\omega} {\tilde {b}} (\frac {1} {\tilde {b}} -1) - 2(1- \omega /4)
\frac {1} {\tilde {b}^2} (\tilde {\rho} + 3\tilde {p}).
 \end{eqnarray}
By using (10) we can transform the torsion function $S_1$ determined by (4) to
the following dimensionless form:
\begin{eqnarray}\label{13}
 \tilde{S}_1 = -\frac{3}{4\tilde{b}Z} (\tilde{H}\tilde{D}+\tilde{E})  ,
\end{eqnarray}
where
\begin{eqnarray}\label{14}
\tilde{D} = \frac{1}{2} \left(3\frac{d \tilde{p}_m}{d \tilde{\rho}_m}-1 \right)
\left(\tilde{\rho}_m+\tilde{p}_m\right)
 +\frac{1}{3}\left(\tilde{\rho}_m- 3\tilde{p}_m\right)+\frac{2}{3}\tilde{\phi}'^2+\frac{4}{3}
 \tilde{V}- \frac{\omega \tilde{b}}{6(1-\omega/4)} \sqrt{X}
 \nonumber\\
+\frac{1-(\omega /2\tilde{b})}{2\sqrt{X}}\Big [\left(3\frac{d
\tilde{p}_m}{d\tilde{\rho}_m}+1 \right)
\left(\tilde{\rho}_m+\tilde{p}_m\right)+ 4 \tilde{\phi}'^2 \Big] + \frac{\omega
(1-\tilde{b}/2)}{3(1-\omega/4)},
\nonumber\\
\tilde{E} = \left(1+ \frac{1-(\omega /2\tilde{b})}{\sqrt{X}}
\right)\frac{\partial \tilde{V}}{\partial \tilde{\phi}}\tilde{\phi}',
Z=\frac{-\omega/4 + (\tilde {b}/2)(1+ \sqrt{X})} {1-\omega/4}.
\end{eqnarray}
Dimensionless form of cosmological equations (2)-(3) obtained by multiplying
these equations on ($6f_0 \omega\alpha$) and by using (11)-(14) is the
following:
\begin{eqnarray}\label{15}
    \frac{k}{\tilde{R}^2} + \Big [\tilde{H}(1+ \frac{3}{2\tilde{b}Z} \tilde{D}) + \frac{3}{2\tilde{b}Z}\tilde{E} \Big]^2=
\frac{1}{Z} \Bigg [\tilde{\rho}  + (1/2) (\frac {Z} {\tilde{b}} -1)
\nonumber\\
\Big [\tilde{\rho} -3\tilde{p} +
   \omega \frac {1-(\tilde{b}/2)(1+
   \sqrt {X})} {1-\omega/4} \Big]
+ \omega \frac{[1-(\tilde{b}/2)(1+\sqrt{X})]^2} {4 (1-\omega/4)^2 } \Bigg],
\end{eqnarray}
\begin{eqnarray}\label{16}
   (\tilde{H}' + \tilde{H}^2)\Big(1+ \frac{3}{2\tilde{b}Z} \tilde{D} \Big) +
\frac{3}{2\tilde{b}Z} \Big [\tilde{H}(\tilde{D}'-\frac{Z'}{Z} \tilde{D}
+\tilde{E}) +E' - \frac{Z'}{Z}\tilde{E} \Big]
\nonumber\\
=-\frac{1}{2Z} \Big[\tilde{\rho} +3\tilde{p}- \omega
\frac{[1-(\tilde{b}/2)(1+\sqrt{X})]^2} {2 (1-\omega/4)^2 } \Big].
\end{eqnarray}
By using the obtained dimensionless form of relations (10)-(16), we will
analyze below the behavior of cosmological solutions for accelerating Universe
at the beginning of cosmological expansion.

\section{Regular Properties of Cosmological Models of Accelerating Universe}

Cosmological equations (15)-(16) contain $\sqrt{X}$, and the condition $X \ge
0$ leads to principal constraint for admissible energy densities and pressures.
In the case of models without scalar field, the equality $X=0$ determines a
limiting energy density, near by which the gravitational interaction is
repulsive \footnote{Since cosmological equations and their solutions are
considered below only in dimensionless form, the "tilde" is omitted in this
Section.}. In the case of models containing also scalar field the equality
$X=0$ determines in space of matter parameters ($\rho_m, \phi, \phi'$) a
limiting $L$-surface ensuring the existence of limiting energy density, which
is different for various solutions. Now we will analyze properties of
cosmological solutions when $X\ll 1$.

Cosmological equation (15) leads to the following expression for the Hubble
parameter:
\begin{eqnarray}\label{17}
H_{\pm}= H_{L} \Bigg [1+ \frac{\sqrt{X}}{(1-(\omega/2b))\frac{\partial
V}{\partial \phi}\phi'} \Big [\frac{\partial V}{\partial \phi}\phi' {\mp}
\nonumber\\
\frac{2bZ}{3} [\frac{1}{Z} (\rho_m + \frac{1}{2} \phi'^2 +V +\frac{1}{2}
(\frac{Z}{b}-1) (\rho_m -3p_m -\phi'^2 +4V
\nonumber\\
+ \omega \frac {1-(b/2)(1+ \sqrt {X})} {1-\omega/4}) + \omega
\frac{(1-(b/2)(1+\sqrt{X}))^2} {4 (1-\omega/4)^2 })
\nonumber\\
-\frac{k}{R^2}]^{1/2} \Big] \Bigg]
\Bigg[1+\frac{2}{(1-\frac{\omega}{2b})[\left(3\frac{d p_m}{d\rho_m}+1 \right)
\left(\rho_m+p_m\right)+ 4 \phi'^2]} \Big [\sqrt{X} (\frac{2bZ}{3}
\nonumber\\
+\frac{1}{2} \left(3\frac{d p_m}{d \rho_m}-1 \right) \left(\rho_m+p_m\right)
+\frac{1}{3}\left(\rho_m- 3p_m\right)+\frac{2}{3}\phi'^2+\frac{4}{3} V+
\nonumber\\
\frac{\omega (1-b/2)}{3(1-\omega/4)})-X \frac{\omega b}{6(1-\omega/4)} \Big]
\Bigg]^{-1},
\end{eqnarray}
where
\begin{eqnarray}\label{18}
H_L=\frac{-2 \frac{\partial V}{\partial \phi}\phi'}{(3\frac{d p_m}{d\rho_m}+1)
\left(\rho_m+p_m\right)+4 \phi'^2}.
\end{eqnarray}
In the case of HIM without scalar field the Hubble parameter vanishes by
reaching a limiting energy density, and $H_{-}$- and $H_{+}$-solutions describe
compression and expansion stage respectively. If $X\ll 1$, the expression (17)
can be written as
\begin{eqnarray}\label{19}
H_{\pm}= H_{L} \frac {1+ B_1 \sqrt {X} +B_2 X +...} {1+ C_1 \sqrt {X} +C_2 X
+...},
\end{eqnarray}
where
\begin{eqnarray}\label{20}
B_1= \frac{1}{(1-(\omega/2b))\frac{\partial V}{\partial \phi}\phi'}\Bigg
[\frac{\partial V}{\partial \phi}\phi' {\mp} \frac{2bZ^{(0)}}{3} \Big
[\frac{1}{Z^{(0)}} (\rho_m + \frac{1}{2} \phi'^2 +V
\nonumber\\
+\frac{1}{2} (\frac{Z^{(0)}}{b}-1) (\rho_m -3p_m -\phi'^2 +4V + \omega \frac
{1-(b/2)} {1-\omega/4}) + \omega \frac{(1-(b/2))^2} {4 (1-\omega/4)^2 })
-\frac{k}{R^2}\Big]^{1/2} \Bigg],
\nonumber\\
B_2={\mp}\frac{2b}{3(1-(\omega/2b))\frac{\partial V}{\partial \phi}\phi'}\Bigg
[Z^{(1)}\Big [\frac{1}{Z^{(0)}} (\rho_m + \frac{1}{2} \phi'^2 +V
\nonumber\\
+\frac{1}{2} (\frac{Z^{(0)}}{b}-1) (\rho_m -3p_m -\phi'^2 +4V + \omega \frac
{1-(b/2)} {1-\omega/4}) + \omega \frac{(1-(b/2))^2} {4 (1-\omega/4)^2 })
-\frac{k}{R^2}\Big]^{1/2}
\nonumber\\
- \Big[\frac{1}{4(1- \frac{\omega}{2b})}(\rho_m +3p_m +2\phi'^2 -2V -\omega
\frac{1-b/2}{1-\omega/4} + \omega \frac{(1-b/2)^2}{(1-\omega/4)^2})-
\nonumber\\
\frac{\omega b(1+\omega/2-b-\omega/2b)}{4(1-\omega/4)^2}\Big] \Big
[\frac{1}{Z^{(0)}} (\rho_m + \frac{1}{2} \phi'^2 +V
\nonumber\\
+\frac{1}{2} (\frac{Z^{(0)}}{b}-1) (\rho_m -3p_m -\phi'^2 +4V + \omega \frac
{1-(b/2)} {1-\omega/4}) + \omega \frac{(1-(b/2))^2} {4 (1-\omega/4)^2 })
-\frac{k}{R^2}\Big]^{-1}\Bigg],
\nonumber\\
C_1=\frac{2}{(1-(\omega/2b))[\left(3\frac{d p_m}{d\rho_m}+1 \right)
\left(\rho_m+p_m\right)+ 4 \phi'^2]}\Big[\frac{2b}{3} Z^{(0)}+
\nonumber\\
\frac{1}{2} \left(3\frac{d p_m}{d \rho_m}-1 \right)
\left(\rho_m+p_m\right)+\frac{1}{3}\left(\rho_m-
3p_m\right)+\frac{2}{3}\phi'^2+\frac{4}{3} V+ \frac{\omega
(1-b/2)}{3(1-\omega/4)}\Big],
\nonumber\\
C_2=\frac{2}{(1-(\omega/2b))[\left(3\frac{d p_m}{d\rho_m}+1 \right)
\left(\rho_m+p_m\right)+ 4 \phi'^2]}\Big[\frac{2b}{3} Z^{(1)}-\frac{\omega
b}{6(1-\omega/4)}\Big],
\end{eqnarray}
and
\begin{eqnarray}\label{21}
Z=Z^{(0)}+ Z^{(1)}\sqrt {X},
\nonumber\\
Z^{(0)}=\frac{b(1-\omega/(2b))}{2(1-\omega/4)},
\nonumber\\
Z^{(1)}=\frac{b}{2(1-\omega/4)}.
\end{eqnarray}
As a result the Hubble parameter can be written in the form of expansion in
$\sqrt {X}$:
\begin{eqnarray}\label{22}
H_{\pm}= H_{L}[1+ (B_1-C_1)\sqrt {X}+(B_2+C_1^2-C_2-B_1C_1) X+...].
\end{eqnarray}
Any characteristics $F$ of HIM can be presented near $L$-surface similarly:
\begin{eqnarray}\label{23}
F_{\pm}= F^{(0)}+ F^{(1/2)}\sqrt {X}+ F^{(1)} X+...,
\end{eqnarray}
where coefficients of expansion $F^{(0)}$, $F^{(1/2)}$, $F^{(1)}$... are some
functions of matter parameters. If some terms of expansion coefficients (23)
contain two signs, we assume that the upper (lower) sign is related to
$H_{+}$-solution ($H_{-}$-solution) respectively. The function $F$ is
continuous on $L$-surface ($X=0$), if $F^{(0)}$ is continuous. Obviously
coefficients of expansion (23) in the case of the Hubble parameter are the
following:
\begin{eqnarray}\label{24}
H^{(0)}=H_{L},
\nonumber\\
H^{(1/2)}=H_{L}(B_1-C_1),
\nonumber\\
H^{(1)}=H_{L}(B_2+C_1^2-C_2-B_1C_1).
\end{eqnarray}
We can obtain the expansion for the torsion function $S_1$ determined by (13)
in the form of expansion (23). It is essential that terms of $D$ and $E$
defined by (14), which are proportional to $(\sqrt {X})^{-1}$, are reduced
mutually by virtue of formula (18). As result we have:
\begin{eqnarray}\label{25}
 S_1 = -\frac{3}{4bZ} (H_{\pm}D+E)  ,
 \nonumber\\
S_1=S_1^{(0)}+S_1^{(1/2)}\sqrt {X}+...,
\nonumber\\
S_1^{(0)}=\frac{1}{2} H_{L}{\mp} \frac{1}{2} \Bigg[\frac{1}{Z^{(0)}}
\Big[\rho_m + \frac{1}{2} \phi'^2 +V
\nonumber\\
+\frac{1}{2} (\frac{Z^{(0)}}{b}-1) (\rho_m -3p_m -\phi'^2 +4V + \omega \frac
{1-(b/2)} {1-\omega/4}) + \omega \frac{(1-(b/2))^2} {4 (1-\omega/4)^2 }\Big]
-\frac{k}{R^2}\Bigg]^{1/2},
\nonumber\\
S_1^{(1/2)}=\frac{1}{1-\omega/2b)}(\frac{\omega}{4b}H_{L}-S_1^{(0)})
-\frac{3}{4bZ^{(0)}}H^{(1/2)}\Big[\frac{1}{2} \left(3\frac{d p_m}{d \rho_m}-1
\right)
\nonumber\\
\left(\rho_m+p_m\right) +\frac{1}{3}\left(\rho_m-
3p_m\right)+\frac{2}{3}\phi'^2+\frac{4}{3} V+ \frac{\omega
(1-b/2)}{3(1-\omega/4)}\Big]-
\nonumber\\
\frac{3(1-\omega/4)}{4b^2}H^{(1)}\Big[\left(3\frac{d p_m}{d\rho_m}+1 \right)
\left(\rho_m+p_m\right)+ 4 \phi'^2\Big],...
\end{eqnarray}
Unlike the Hubble parameter (22), which is continuous function at $L$-surface,
the torsion function $S_1$ according to (25) undergoes a finite jump by
transition from $H_{-}$- to $H_{+}$-solution. Though the expansion (23) for
$F_{\pm}$ contains a term with $\sqrt {X}$, the expansion for derivative
$F_{\pm}'$ is regular because the derivative of $\sqrt {X}$ does not diverge at
$X=0$ by virtue of (18), namely we have:
\begin{eqnarray}\label{26}
\frac{X'}{\sqrt {X}}=\frac{6}{b^2}(1-\omega/4)(H^{(1/2)}+H^{(1)}\sqrt {X}+...)
\nonumber\\
\Big[\left(3\frac{d p_m}{d\rho_m}+1 \right) \left(\rho_m+p_m\right)+ 4 \phi'^2
\Big].
\end{eqnarray}
As result the derivative of the Hubble parameter near $L$-surface is:
\begin{eqnarray}\label{27}
H_{\pm}'=H_{L}'+ \frac{3}{b^2}(1+\omega/4)H^{(1/2)2}\Big[\left(3\frac{d
p_m}{d\rho_m}+1 \right) \left(\rho_m+p_m\right)+ 4 \phi'^2
\Big]+\Bigg[H^{(1/2)'}+
\nonumber\\
\frac{9}{b^2}(1-\omega/4)H^{(1/2)}H^{(1)}\Big[\left(3\frac{d p_m}{d\rho_m}+1
\right) \left(\rho_m+p_m\right)+ 4 \phi'^2 \Big]\Bigg]\sqrt {X}+...
\end{eqnarray}
and for derivative $S_1'$ we obtain the following expansion:
\begin{eqnarray}\label{28}
S_1'=S_1^{'(0)}+S_1^{'(1/2)}\sqrt {X}+...,
\nonumber\\
S_1^{'(0)}=S_1^{(0)'}+\frac{3}{b^2}(1-\omega/4)H^{(1/2)}\Big[\left(3\frac{d
p_m}{d\rho_m}+1 \right) \left(\rho_m+p_m\right)+ 4 \phi'^2\Big]S_1^{(1/2)},...
\end{eqnarray}

We see that from mathematical point of view the limiting $L$-surface plays a
special role because some physical characteristics undergo a finite jump by
transition from $H_{-}$- to $H_{+}$-solution. However, it is of principal
meaning that all physical characteristics including the torsion functions and
curvature tensor are regular for $H_{-}$- and $H_{+}$-solutions and do not
diverge by approaching to limiting $L$-surface. As it follows from our analysis
the Hubble parameter, the torsion function $S_2$ and the curvature functions
$A_1$ and $A_2$ determining the structure of cosmological equations (2)-(3) are
continuous on limiting $L$-surface while the torsion function $S_1$, the
derivative $S_2'$ and as result the curvature functions $A_3$ and $A_4$ undergo
a finite jump by transition from $H_{-}$- to $H_{+}$-solution. In the case of
HIM without scalar field the limiting $L$-surface is transformed into state
with limiting energy density ($X=0$) which corresponds to a bounce, and in this
case according to obtained formulas (27)-(28) the derivative of the Hubble
parameter and derivative $S_1'$ are continuous by transition from $H_{-}$- to
$H_{+}$-solution. In the case of HIM with scalar field a bounce takes place in
points of extremum surface, which we obtain from cosmological equation (15) by
supposing that the Hubble parameter vanishes. In this case all physical
characteristics of HIM including inflationary models are continuous at a
bounce. Cosmological solutions for inflationary HIM can be found by numerical
integration of eqs. (16) and (10) by choosing initial conditions for $(\rho_m,
\phi, \phi')$ on extremum surface $H=0$ \cite{a15}.

As it is known, the PGTG based on gravitational Lagrangian (1) by certain
restrictions on indefinite parameters of ${\cal L}_{\rm g}$ is free of such
pathological objects as ghosts and tachyons that was shown by investigation of
gravitational perturbations in Minkowski and Einsteinian backgrounds
\cite{a6,a16,a17}. Restrictions on indefinite parameters used in this paper and
allowed the regularity of all HIM are compatible with corresponding
restrictions excluding ghosts and tachyons which are obtained in Ref. 6:
conditions $a=0$ and $q_2=0$ lead to excluding of massive particles with
spin-parity $0^+$ and $0^-$ respectively, and the conditions $\omega \alpha>0$,
which ensures the appearance of limiting energy density, is compatible with the
presence of particles with spin-parity $2^-$ and $1^-$ by excluding particles
$2^+$ and $1^+$ \footnote{It is particular case of the class I introduced in
\cite{a6} of particle content in PGTG.}. As it was noted in Ref. 13, if the
physical spacetime in the vacuum has the structure of Riemann-Cartan continuum
with de Sitter metrics and non-vanishing torsion, the particle content of PGTG
has to be investigated on such background. However, this hard problem is not
yet solved. It should be noted that the change of the structure of vacuum
spacetime in the frame of PGTG leads to principal differences of gravitational
interaction in comparison with other fundamental physical interactions (by
supposing that PGTG is correct gravitation theory), and possibly the search of
gravitational interaction requires not traditional approach.

\section{Conclusion}

The investigation of isotropic cosmology built in the framework of PGTG
presented above shows that this theory allows one to solve the PCS on the base
of classical consideration in four-dimensional physical spacetime. Unlike HIM
with the only torsion function $S_1$ \cite{a7,a8}, in the case of which the
torsion and hence the curvature diverge by transition from $H_{-}$- to
$H_{+}$-solution, all HIM with two torsion functions (by certain restrictions
on indefinite parameters) are regular not only with respect to metrics with its
time derivatives and energy density, but also with respect to the torsion and
curvature tensors. Their regular behaviour is of principal meaning for
consistent description of HIM in the frame of classical theory. The
investigation of applicability limits of obtained physical results in the case
of gravitating systems with lesser spacial symmetry is of principal interest
for cosmology as well as for astrophysics. In particular, the existence of
limiting energy density and gravitational repulsion effect at extreme
conditions can explain the presence of massive objects in galaxies centrum.

\end{document}